# In-vivo measurement of the human soft tissues constitutive laws. Applications to Computer Aided Surgery.

Schiavone P.[1,2], Boudou T.[1], Ohayon J.[1], Payan Y.[1]

[1] Laboratoire TIMC-IMAG, UMR UJF CNRS 5525, La Tronche, France
[2] Laboratoire des Technologies de la Microélectronique, CNRS, Grenoble, France

**Introduction**

In the 80's, biomechanicians were asked to work on Computer Aided Surgery applications since orthopaedic surgeons were looking for numerical tools able to predict risks of fractures. More recently, biomechanicians started to address soft tissues arguing that most of the human body is made of such tissues that can move as well as deform during surgical gestures (Payan, 2005). An intra-operative use of a continuous Finite Element (FE) Model of a given tissue mainly faces two problems: (1) the numerical simulations have to be "interactive", i.e. sufficiently fast to provide results during surgery (which can be a strong issue in the context of hyperelastic models for example) and (2) during the intervention, the surgeon needs a device that can be used to provide to the model an estimation of the patient-specific constitutive behaviour of the soft tissues.

This work proposes an answer to the second point, with the design of a new aspiration device aiming at characterizing the in vivo constitutive laws of human soft tissues. The device was defined in order to permit sterilization as well an easy intra-operative use.

**Materials and methods**

Aspiration experiments have proved to be an efficient way for measuring soft tissue mechanical properties (Nava, 2003). The very severe constraints induced by the per-operative use targeted by our device lead to the conception of the simplest possible setup. Especially a way for solving sterilization issues was to avoid any electronic part in the device. A small depression (few tens of mbars) is generated inside a transparent cylinder in contact with the material in test. The bump induced by the depression is imaged from the top using a standard digital camera in macro mode thanks to a 45° mirror (Figure 1).

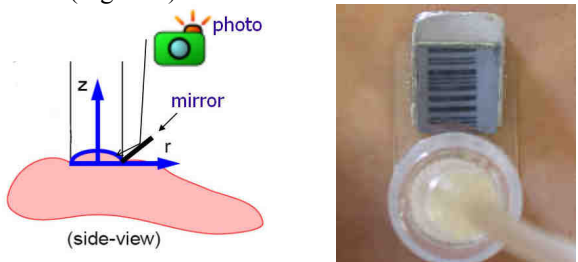

Figure 1: pipette aspiration experiment

In order to estimate the tissue constitutive law, an FE model of the aspiration experiment is used to fit the data. The tissue sample is represented by a thick circular slice of radius $a$ and thickness $h$. The pipette is described by a rigid hollow cylinder of internal and external radii $R_i$ and $R_e$ respectively. Taking advantage of the axi-symmetric geometry of the problem, we reduced its mechanical study to a two-dimensional structural analysis.

**Results**

The device was tested on human forearm skin (fig. 1, right). For this tissue, a non-linear response was assumed. A two-parameter Mooney-Rivlin strain energy function was assumed, with $W = a_1(I_1 - 3) + a_3(I_1 - 3)^3$, where $a_1$ and $a_3$ are material constants, while $I_1$ is the first invariant of the Cauchy strain tensor.

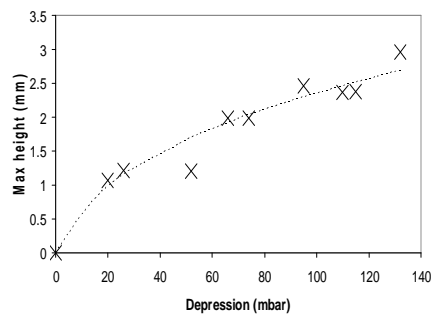

Figure 2 : Finite element simulation compared to measured data

The experimental results are quite well reproduced by the FE simulation (Figure 2). The method consisted in minimizing the difference between measured and simulated sample aspirated lengths for successively imposed negative pressure. The optimal values of the material constants $a_1$ and $a_3$ were then respectively identified to 2 and 400.

## Conclusion

The aspiration technique appears to be a robust measurement method suited for an intra-operative use. Coupled with an optimization scheme based on FEM, an inversion procedure allows determining non-linear materials models. The inverse process which is currently too long can be improved significantly to allow an "interactive time" use.